\input harvmac
\def\npb#1(#2)#3{{ Nucl. Phys. }{B#1} (#2) #3}
\def\plb#1(#2)#3{{ Phys. Lett. }{#1B} (#2) #3}
\def\pla#1(#2)#3{{ Phys. Lett. }{#1A} (#2) #3}
\def\prl#1(#2)#3{{ Phys. Rev. Lett. }{#1} (#2) #3}
\def\mpla#1(#2)#3{{ Mod. Phys. Lett. }{A#1} (#2) #3}
\def\ijmpa#1(#2)#3{{ Int. J. Mod. Phys. }{A#1} (#2) #3}
\def\cmp#1(#2)#3{{ Commun. Math. Phys. }{#1} (#2) #3}
\def\cqg#1(#2)#3{{ Class. Quantum Grav. }{#1} (#2) #3}
\def\jmp#1(#2)#3{{ J. Math. Phys. }{#1} (#2) #3}
\def\anp#1(#2)#3{{ Ann. Phys. }{#1} (#2) #3}
\def\prd#1(#2)#3{{ Phys. Rev.} {D\bf{#1}} (#2) #3}

\def\inbar{\,\vrule height1.5ex width.4pt depth0pt}
\def\IQ{\relax\,\hbox{$\inbar\kern-.3em{\rm Q}$}}
\def\IB{\relax{\rm I\kern-.18em B}}
\def\IC{\relax\hbox{$\inbar\kern-.3em{\rm C}$}}
\def\IP{\relax{\rm I\kern-.18em P}}
\def\IR{\relax{\rm I\kern-.18em R}}
\def\ZZ{\relax\ifmmode\mathchoice
{\hbox{Z\kern-.4em Z}}{\hbox{Z\kern-.4em Z}}
{\lower.9pt\hbox{Z\kern-.4em Z}}
{\lower1.2pt\hbox{Z\kern-.4em Z}}\else{Z\kern-.4em Z}\fi}
\def\hb{\hfill\break}

\def\n*#1{\nu^{* (#1)}}

\def\IP{{\bf P}}

\def\n#1{\nu_{#1}^*}

\def\IP{{\bf P}}

\def\n#1{\nu_{#1}^*}

\def\({ \left(  }
\def\){ \right) }

\def\ph{\phantom- }

\def\-{\phantom{-}}
\noblackbox

\def\cicy#1(#2|#3)#4{\left(\matrix{#2}\right|\!\!
                     \left|\matrix{#3}\right)^{{#4}}_{#1}}

\catcode`@=12
\baselineskip16pt
\noblackbox
\newif\ifdraft

\noblackbox
\newif\ifhypertex
\ifx\hyperdef\UnDeFiNeD
    \hypertexfalse
    \message{[HYPERTEX MODE OFF}
    \def\hyperref#1#2#3#4{#4}
    \def\hyperdef#1#2#3#4{#4}
    \def\hypernoname{}
    \def\e@tf@ur#1{}
    \def\hth/#1#2#3#4#5#6#7{{\tt hep-th/#1#2#3#4#5#6#7}}
    
\else
    \hypertextrue
    \message{[HYPERTEX MODE ON}
  \def\hth/#1#2#3#4#5#6#7{
  {\tt hep-th/#1#2#3#4#5#6#7}}

\fi

\catcode`\@=11
\newif\iffigureexists
\newif\ifepsfloaded
\def\epsfcheck{
\ifdraft
\input epsf\epsfloadedtrue
\else
  \openin 1 epsf
  \ifeof 1 \epsfloadedfalse \else \epsfloadedtrue \fi
  \closein 1
  \ifepsfloaded
    \input epsf
  \else
\immediate\write20{NO EPSF FILE --- FIGURES WILL BE IGNORED}
  \fi
\fi
\def\epsfcheck{}}
\def\checkex#1{
\ifdraft
\figureexistsfalse\immediate%
\write20{Draftmode: figure #1 not included}
\else\relax
    \ifepsfloaded \openin 1 #1
        \ifeof 1
           \figureexistsfalse
  \immediate\write20{FIGURE FILE #1 NOT FOUND}
        \else \figureexiststrue
        \fi \closein 1
    \else \figureexistsfalse
    \fi
\fi}
\def\missbox#1#2{$\vcenter{\hrule
\hbox{\vrule height#1\kern1.truein
\raise.5truein\hbox{#2} \kern1.truein \vrule} \hrule}$}
\def\lfig#1{
\let\labelflag=#1%
\def\numb@rone{#1}%
\ifx\labelflag\UnDeFiNeD%
{\xdef#1{\the\figno}%
\writedef{#1\leftbracket{\the\figno}}%
\global\advance\figno by1%
}\fi{\hyperref{}{figure}{{\numb@rone}}{Fig.{\numb@rone}}}}
\def\figinsert#1#2#3#4{
\epsfcheck\checkex{#4}%
\def\figsize{#3}%
\let\flag=#1\ifx\flag\UnDeFiNeD
{\xdef#1{\the\figno}%
\writedef{#1\leftbracket{\the\figno}}%
\global\advance\figno by1%
}\fi
\goodbreak\midinsert%
\iffigureexists
\centerline{\epsfysize\figsize\epsfbox{#4}}%
\else%
\vskip.05truein
  \ifepsfloaded
  \ifdraft
  \centerline{\missbox\figsize{Draftmode: #4 not included}}%
  \else
  \centerline{\missbox\figsize{#4 not found}}
  \fi
  \else
  \centerline{\missbox\figsize{epsf.tex not found}}
  \fi
\vskip.05truein
\fi%
{\smallskip%
\leftskip 4pc \rightskip 4pc%
\noindent\ninepoint\sl \baselineskip=11pt%
{\bf{\hyperdef\hypernoname{figure}{{#1}}{Fig.{#1}}}:~}#2%
\smallskip}\bigskip\endinsert%
}
%


\nopagenumbers\abstractfont\hsize=\hstitle
\null
\rightline{\vbox{\baselineskip12pt\hbox{CERN-TH/96-133}
                                  \hbox{NSF-ITP-95-162}
                                  \hbox{OSU-M-96-6}
                                  \hbox{hep-th/9605154}}}%
\vfill
\centerline{\titlefont New Higgs Transitions between Dual N=2 String Models}
\vskip5pt
\abstractfont\vfill\pageno=0

\vskip-0.8cm
\centerline{Per Berglund}                                \vskip-.2ex
 \centerline{\it Theory Division, CERN}      \vskip-.4ex
 \centerline{\it CH-1211 Geneva  23, Switzerland}       \vskip-.4ex
\vskip .05in
\centerline{Sheldon Katz}                                \vskip-.2ex
 \centerline{\it Department of Mathematics} \vskip-.4ex
 \centerline{\it Oklahoma State University}                   \vskip-.4ex
 \centerline{\it Stillwater, OK 74078, USA} \vskip0ex
\vskip .05in
\centerline{Albrecht Klemm}                                \vskip-.2ex
 \centerline{\it Department of Mathematics, Harvard University} \vskip-.4ex
 \centerline{\it Cambridge MA 02138, USA}\vskip-.4ex
\vskip .05in
\centerline{Peter Mayr\footnote{$^{}$}
      {Email: berglund@nxth04.cern.ch, katz@math.okstate.edu,
klemm@abel.math.harvard.edu, mayr@surya11.cern.ch}}       \vskip-.2ex
 \centerline{\it Theory Division, CERN}      \vskip-.4ex
 \centerline{\it CH-1211 Geneva  23, Switzerland}       \vskip-.4ex
\vskip .05in
\vfill
\vskip-0.3cm
\vbox{\narrower\baselineskip=12pt\noindent
We describe a new kind of transition between topologically distinct
$N=2$ type II Calabi--Yau vacua through points with enhanced non-abelian
gauge symmetries together with fundamental charged matter 
hyper multiplets. We connect the appearance of matter to the 
local geometry of the singularity and discuss the relation between the
instanton numbers of the Calabi--Yau manifolds taking part in the
transition. In a dual heterotic string theory on $K3\times T^2$ the
process corresponds to Higgsing a semi-classical gauge group or
equivalently to a variation of the gauge bundle.
In special cases the situation reduces to simple conifold 
transitions in  the Coulomb phase of the non-abelian gauge symmetries.}

\Date{\vbox{\line{CERN-TH/96-133\hfill}
            \line{5/96 \hfill}}}

\vfill\eject
\baselineskip=14pt plus 1 pt minus 1 pt

\newsec{Introduction}
During the last few years there has been a great deal of progress in
the understanding of non-perturbative phenomena in supersymmetric field 
theories as well as in various string theories
\ref\witten1{E. Witten, 
\npb{443} (1995) 85, hep-th/9503124.}. In particular the idea of duality has 
proven to be crucial. The basic point here is that one underlying 
theory might have several descriptions in terms of physical variables.
In the current setting these ideas originated in the work by Seiberg 
and Witten in the context of $N=2$ supersymmetric Yang-Mills 
theory~\ref\sw{N. Seiberg, E. Witten,
\npb{426}(1994) 19, 
hep-th/9407087.}.  For all gauge groups the global prepotential 
can be derived from the periods of a suitable auxiliary 
Riemann-surface~\ref\othergauge{
A. Klemm, W. Lerche, S. Theisen and S. Yankielowicz,
Phys. Lett. 344 (1995) 169,
P. Argyres and A. Faraggi, 
Phys. Rev. Lett. 74  (1995) 3931; 
Ulf H. Danielsson, Bo Sundborg, \plb358(1995)273;  
A. Brandhuber, K. Landsteiner, Phys.Lett.B358 (1995) 73, 
R.G. Leigh, M.J. Strassler, Phys.Lett.B 356 (1995) 492; 
E. Martinec, N. Warner,\npb459(1996)97}.   
Combining the properties of $N=2$ gauge theories with the conjectured 
duality between $N=4$ string theories in four dimensions 
\ref\huto{see e.g. C. Hull and P. Townsend, \npb{438} (1995) 109, 
hep-th 9410167}  it was  suggested that also $N=2$ string 
theories in four dimensions have a duality structure; 
the heterotic string theory compactified on 
$K3\times T^2$ and the type II theory compactified on a Calabi-Yau manifold, 
could be dual to each other after including non-perturbative 
states~\ref\KaVa{S. Kachru and C. Vafa,
\npb{450} (1995) 69, hep-th/9505105.}\ref\FHSV{S. Ferrara, J. Harvey, 
A. Strominger, and 
C. Vafa {\sl Second-Quantized Mirror Symmetry}, hep-th/9505162}. 
 
In~\KaVa\ concrete pairs of dual $N=2$ theories were constructed
in which non-perturbative properties of the heterotic 
string can be investigated exactly. The key idea is the absence of 
neutral couplings between vector multiplets and hyper multiplets in 
$N=2$ theories~\ref\specialkaehler{B. de Wit, P.G. Lauwers, 
R. Philippe, S.Q. Su, A. van Proeyen, \plb134(1984)37; 
B. de Wit, A. van Proeyen, \npb245(1984)89; J.P. Derendinger, 
S. Ferrara, A. Masiero,
A. van Proeyen, \plb140(1984)307; B. de Wit, P.G. Lauwers,
A. van Proeyen, \npb255(1985)569; E. Cremmer, C. Kounnas, A. van Proeyen,
J.P. Derendinger, S. Ferrara, B. de Wit, L. Girardello,
\npb250(1985)385}. As the heterotic dilaton sits in a vector 
multiplet the vector multiplet moduli space can receive space-time 
perturbative and non-perturbative corrections. 
Under $N=2$ string-string-duality it is identified with the vector 
multiplet moduli space of the Type IIA string, which corresponds to 
the K\"ahler moduli space of the compactified Calabi-Yau manifold.
The latter has to have Hodge numbers 
$h_{11}=N_v$ and 
$h_{21}=N_h-1$, where $N_v$, $N_h$ are the number of vector 
and hyper multiplets and the $-1$ corresponds to the 
the dilaton of the type II string. (Note however, that the rank of the
gauge group is $N_v+1$ as the graviphoton contributes a $U(1)$-factor.)
As the Type IIA dilaton is in a hyper multiplet the vector multiplet
moduli space is exact at tree level and so is not corrected by 
space-time instantons. It is, however, corrected by worldsheet instantons. 
By mirror symmetry we can identify it with the hyper multiplet moduli space 
of the Type IIB theory on the mirror Calabi-Yau manifold, 
which receives neither world-sheet nor space-time corrections. 
The upshot is that the exact non-perturbative vector multiplet 
moduli space of the heterotic string is modelled by the complex 
structure moduli space of a specific Calabi-Yau 
manifold~\foot{Similarly, the structure of the non-perturbative hyper 
moduli space 
of the type IIA theory, which is a quaternionic 
manifold~\ref\bw{J. Bagger,
E. Witten, \npb222(1983)1;
J. Bagger, A. Galperin. E. Ivanov, 
V. Ogievetsky, \npb303(1988)522} can be investigated via the heterotic 
string.}. In this sense the complex moduli space of Calabi-Yau manifold 
replaces the complex moduli space of the auxiliary Riemann 
surface, which one had in the $N=2$ Yang-Mills field theory case.

Following the initial work~\KaVa\FHSV\ a number of consistency 
checks have been made further establishing  the conjectured
 type II/heterotic string duality; comparison of the 
perturbative region of the potentials~\ref\pert{
B.\ de Wit, V.\ Kaplunovsky, J.\ Louis and D.\ L\"ust, \npb451(1990) 53;
I.\ Antoniadis, S.\ Ferrara, E.\ Gava, K.\ S.\ Narain 
and T.\ R.\ Taylor, \npb447 (1995) 35}
of the heterotic couplings (gauge and 
gravitational) with that of the dual type IIA vacuum 
\KaVa, \ref\ogen{
V.\ Kaplunovsky, J.\ Louis, and S.\ Theisen, \plb357 (1995) 71;
A.\ Klemm, W.\ Lerche and P.\ Mayr, \plb357 (1995) 313;
C.\ Vafa and E.\ Witten, preprint HUTP-95-A023;
hep-th/9507050; I.\ Antoniadis, E.\ Gava, K.\ Narain and 
T.\ Taylor, \npb455 (1995) 109,
B. Lian and S.T. Yau,
{\sl Mirror Maps, Modular Relations and Hypergeometric Series I,II},
hep-th/9507151, hep-th 9507153;P.\ Aspinwall and J.\ Louis, \plb369 (1996) 233;
I.\ Antoniadis, S.\ Ferrara and T.\ Taylor,\npb 460(1996)489;
G. Curio, \plb366 (1996) 131,\plb366 (1996) 78;
G.\ Lopes Cardoso, G. Curio, D. L\"ust and T. Mohaupt,
{\sl Instanton Numbers and Exchange Symmetries in $N=2$ Dual String
Pairs}, hep-th/9603108}, as
well as non-perturbative consistency checks
in the 
point-particle limit~\ref\KKLMV{S. Kachru, A. Klemm, W. Lerche, P. Mayr and C. 
Vafa, \npb459(1996)537}.

Furthermore, and what will be the focus of this paper,
aspects of the perturbative enhancements of the gauge symmetry 
can be studied by considering the Picard lattice of the generic 
$K_3$-fiber~\ref\AspinGauge{P. Aspinwall, \plb{357} (1995) 329, hep-th/9507012 
\semi {\sl Enhanced Gauge Symmetries and Calabi-Yau Threefolds}, 
hep-th/9511171.}. Using
the Higgs mechanism as a way of lowering the rank of the gauge group 
and thus finding a way in which the various moduli spaces can 
be connected\foot{This so called Higgs-branch has to be 
distinguished from the Higgs breaking mechanism into the Coulomb-branch
by vector multiplets, in which the gauge bosons become massive as
short vector multiplets under spontaneous generation of central charge,
as in the Seiberg-Witten theory.}\ref\connect{A. C. Avram, P.\  Candelas,
D.\ Jancic, M. Mandelberg, {\sl On the connectedness 
of moduli spaces of Calabi-Yau manifolds } hep-th/9511230;
T, Chiang, B. R. \ Greene, M. Gross, Y. Kanter, {\sl
Black hole condensation and the web of Calabi-Yau
manifolds.} hep-th 9511204},
chains of such 
transitions have been studied 
extensively~\KaVa\ref\afiq{G. Aldazabal , L.E. Ibanez , 
A. Font , F. Quevedo, {\sl Chains of N=2, D=4 heterotic/type II
duals}, hep-th/9510093.}~\ref\cf{P. 
Candelas and A. Font, {\sl Duality Between Webs of Heterotic and Type 
II Vacua}, hep-th/9603170.}.

Here we will focus on a particular chain and investigate 
in detail the transition from the point of view of the local geometry
as well as in terms of the the worldsheet instanton
sums in the type IIA theory. 
In particular, we will not only be able to identify the enhanced 
perturbative gauge symmetry
but also the matter representations. 
The geometrical transition is of a slightly generalized 
type compared to the conifold transitions on one 
hand and the strong coupling transitions on the other hand. In addition to 
having N-1 divisors being contracted to a singular curve $C$, of genus zero, 
giving rise to an SU(N) theory with no adjoint hyper 
multiplets~\ref\kmp{ S. Katz , D. R. Morrison , 
M. R. Plesser, {\sl Enhanced Gauge Symmetry in Type II String Theory}, 
hep-th/9601108.}, 
there are 
singular fibers, which when contracted to $C$ give rise to  
massless hyper multiplets in the fundamental representation. 
The general structure of these transitions is discussed in section~2. We then
turn to a specific set of models in section~3.
This chain of Calabi-Yau manifolds can be viewed 
as $K_3$ fibrations over $\IP^1$ as well as elliptic fibrations over 
the Hirzebruch surface $F_2$. 
Finally, we end with conclusions and discussions in section~4.

\newsec{Structure of the extremal transitions}
\subsec{Physical spectrum and flat directions}
It will be useful to consider the purely field theoretic problem of which
massless particles will appear in the moduli space. At a generic point 
in the moduli space of vacuum expectation values of the scalar component
of the vector multiplet, the gauge symmetry is $U(1)^{N-1}$. By a suitable
gauge transformation we can then write the above scalar in the
diagonal form $diag(\phi_1,...,\phi_N)$, where $\sum_{i=1}^N \phi_i=0$ and 
$\phi_i\neq \phi_j$, for all $i,j$. Clearly, all the matter fields are
massive with masses proportional to $\phi_i$. If we now set $\phi_i=0,\,
i=1,...,N-1$, the gauge symmetry is enhanced to $SU(N)$ and only
matter fields charged with respect to the $SU(N)$ become massless.
Let us for simplicity assume that there are $M$  fields which
all transform in the fundamental representation. Because of the
tracelessness condition there exists a surface for which one can 
have $SU(N-1)\times U(1)$, with $\phi_i=\phi_j\neq 0,\, i,j=1,...,N-1$ 
but without massless
charged matter. If we reduce the symmetry enhancement one step further there
is however room for massless matter in the fundamental of $SU(N-2)$; e.g.
choose $\phi_i=0,\, i=1,...,N-2$ and $\phi_{N-1}=-\phi_N\neq 0$. In general
one will have several $SU(k_j)$; however, only the one for which 
$\phi_i=0, i=1,...,k_j$ will have $M$
massless matter multiplets. In particular, there exists a codimension one
surface in which only e.g. $\phi_1=0$, such that there is no gauge 
enhancement. However, because of $\phi_1=0$ we will have $M$ massless
singlets. It is very gratifying that, as will now be seen, 
the Calabi-Yau moduli space exactly reproduces this kind of behaviour.

\subsec{Local geometry of the Calabi--Yau singularity}
During the recent developments it has become clear that the
physical singularities associated to massless solitonic 
BPS states are essentially encoded in the geometry
of the singularity of the compactified manifold.
The role played by the geometry can be understood from 
the interpretation of the massless states as solitonic p-branes
wrapped around the vanishing cycles of the
singularity; the gauge and Lorentz quantum numbers depend
then on some characteristic properties of the homology
cycles, in particular their dimension, topology and intersection
numbers~\kmp.

The simplest case (in the type IIB picture) is that of a vanishing
three-cycle leading to a massless hyper multiplet, the case considered
originally by Strominger~\ref\andy{A. Strominger, {\sl Massless Black Holes 
and Conifolds in String Theory }, hep-th/9504090.}. On the other hand, 
if the three-cycle shrinks
to a curve, rather than to a point, one obtains enhanced gauge symmetries,
as has been argued in~\ref\bsv{M. Bershadsky , V. Sadov , C. Vafa, {\sl D-Strings on 
D-Manifolds}, hep-th/9510225.}. More precisely, 
if the local geometry is that of an ALE space with $A_N$ singularity
over a curve of genus $g$ one obtains an  enhanced  $SU(N+1)$ gauge
symmetry~\ref\km{A. Klemm and P. Mayr,
{\sl Strong Coupling Singularities and Non-abelian Gauge Symmetries in 
$N=2$ String Theory}, hep-th/9601014.}\ together with $g$ hyper multiplets in the adjoint 
representation~\kmp.
The case $g=0$ is exceptional in that the enhanced gauge symmetry is
asymptotically free and broken to its abelian factor due to
strong coupling effects in the infrared; this case has been
considered in~\KKLMV~\ref\KLMVW{A. Klemm, W. Lerche, P. Mayr, C. Vafa and N. Warner,
{\sl Selfdual Strings and $N=2$ Supersymmetric Field Theory}, hep-th/9604034}.

Let us now assume that we have a collection of curves $C_i$ with the transverse
space that of an ALE-manifold with $A_{N_i}$ type singularity and 
consider further a point of intersection between two of these 
curves~\ref\mp{D. R. Morrison and M. R. Plesser, private communication.}.
The singularity structure
of the transverse space has been analyzed in detail in~\ref\mir{R.\ Miranda,
in {\sl The Birational Geometry of Degenerations\/}, Progress in Math.\ vol.\ 
29, Birkh\"auser, 19983, 85.}
in the context of elliptic fibrations with the
following result: if along the curves $C_1$ and $C_2$ the elliptic fiber 
is of Kodaira 
type $I_{N_1}$ and $I_{N_2}$ respectively, then above the point where $C_1$
intersects $C_2$ the elliptic fiber is of type 
type $I_{N_1+N_2}$. Indeed the examples we will consider
are all ellipticly fibered Calabi--Yau manifolds.
This allows in particular for a simple interpretation in terms of 
5-branes ~\bsv~\kmp located at the points where the fibration becomes 
singular. However this special structure is not necessary; 
the general configuration is that of a collection of curves $C_i$
with transverse $A_{N_i}$ singularities colliding in a set of $M$
points over which
the singularity structure jumps to $A_{N_i+N_j}$.

A simple D-brane arrangement based on a collection of $N_i$ coinciding
D-branes intersecting a second collection of $N_j$ coinciding 
D-branes has been given in~\bsv. In this picture additional
matter in the fundamental representation 
arises from open string states with one end attached to the
first and the other end to the second collection. There is a special
configuration in which one of the two collections consists
of a single D-brane only, say $N_j=1$. In this case one expects a 
single non-abelian factor of $SU(N_i)$ together with matter in the
fundamental representation.

This is the physical situation whose realization we will consider
in the context of Calabi--Yau compactifications.
Specifically the case of $SU(N+1)$ gauge symmetry with $M$ fundamental
matter multiplets arises from the following local data:
there is a curve $C$, which in our case will be a $P^1$, over
which one has a bundle structure where the generic fiber is a
Hirzebruch-Jung tree of the resolution of the $A_N$ singularity,
that is a collection of $P^1$'s, $E_i,\ i=1..N$  
with intersection matrix proportional
to the Cartan matrix of $A_N$. In addition, above the 
$M$ exceptional points, the singularity structure of the fiber becomes 
$A_{N+1}$ because one component of the fiber factorizes. More precisely out of
the $N$ generic components $N-1$ are toric divisors $D_i$
of the manifold $X_i$ which are ruled surfaces, 
\def\eh{\hat{E}}
while the $N$-th component, $\eh$ is only birationally ruled, 
having $M$ degenerate fibers. The last component $\eh$
is actually a {\it conic bundle\/}, which means that the fibers are all 
plane conics~\ref\joe{J. Harris, {\sl Algebraic Geomtry},
Springer (1992) New York}. These conics are smooth over a generic points
of $C$ while they split into line pairs over the $M$ exceptional points.

Let us describe now how the appearance of this
structure will lead to geometrical transitions between two 
Calabi--Yau manifolds $X_i$ and $X_{i-1}$ (where $i$ denotes
the number of $N_V=h_{11}$ of vector multiplets).
First we can contract the $N-1$ divisors together with the $M$
degenerate fibers of the conic bundle. According to our
previous discussion, the degenerate fibers contain a collection of 
$N+1$ rational curves $E_i$ with intersection matrix of $A_{N+1}$
which are all contracted. $N$ of them are again associated to
the Cartan subalgebra of the gauge group while the additional
one is related to the matter hyper multiplets; there is a natural
action of the $A_N$ Weyl group on this class which generates
the components of the fundamental representation. The gauge quantum
numbers of the solitonic p-brane states are determined by the
reduction of the Cartan matrix of $A_{N+1}$ to that of $A_N$;
the components of the matter fields arise naturally from wrappings
of the cycles $E_{N+1}-E_i$, $i=0,\ldots,N$; however there is no independent
modulus associated to the volume of $E_{N+1}$ and correspondingly 
no additional vector multiplet\foot{
Here $E_0$ is the homology cycle which fulfills the relation
$\sum_0^N E_i \sim 0$, reflecting the tracelessness condition of $SU(N+1)$,
as discussed in the previous section.}.
This is expected from the fact that
the additional rational curves are isolated rather than being
fibers of a ruled surface.
After the $N$ surfaces have been contracted to the base the
singularities can be simplified by a deformation
of the complex structure in such a way that the resulting singularity is the 
contraction of $N-1$ surfaces arising from $X_{i-1}$. This completes the 
transition from $X_i$ to $X_{i-1}$ at the enhanced symmetry points.

Clearly, in the generic situation one will only contract subsets 
$S_i$ of the $N-1$ divisors and/or the conic bundle with the $M$ degenerate 
fibers.
The result will depend on whether a given subset $S_i$ contains $\eh$;
if it does we will get from that factor an enhanced gauge symmetry $SU(k_i+1)$
together with $M$
fundamental representations; however if it does not, the result
is that of a non-abelian gauge symmetry without matter, broken down to the
Cartan subalgebra
by strong infrared dynamics . In fact if
$N>1$ we can contract {\it only} $\eh$ and the result is $M$ representations
of $U(1)$, that is we are back to the familiar case of the conifold 
singularity.

\subsec{String moduli space}
\def\cM{{\cal M}}
We turn now to a discussion of the string moduli spaces 
involved in the transition. The latter is described by a motion 
in the vector multiplet moduli \cM\ space of the Calabi--Yau manifold
$X_i$ to a locus where the vev of the scalar superpartner of a vector field
vanishes and then turning on 
vevs in the new flat directions of the Higgs branch corresponding to a 
motion in the
hyper multiplet moduli space of the Calabi--Yau manifold $X_{i-1}$;
the new Calabi-Yau manifold $X_{i-1}$ will therefore have fewer vector 
moduli while the  number of hyper moduli has increased; 
the associated change in the Hodge numbers $h_{11}$ and $h_{12}$ indicates
that the two manifolds are of different topological type.
There are two types of natural coordinates on $\cM$, the
algebraic coordinates  $z_n$ and the special coordinates $t_n$ 
~\ref\morIII{P. Aspinwall, B. Greene and 
D. Morrison, \npb{420} (1994) 184},
where the two are 
related by the mirrormaps $z_m(t_n)$. We are 
interested in the relations between these two types of coordinates 
on $\cM^i$ and $\cM^{i-1}$.

$N=2$ supersymmetry puts strong restrictions on these relations;
in particular the special geometry of the vector multiplets in 
the type IIA compactification constrains the map between the
two set of coordinates $t_n^i$ and $t^{i-1}_n$ to be linear.
Moreover we will find simple relations
between the Gromov--Witten invariants $n_{i_1,\dots,i_{h_{11}}}$
on $\cM^i$ and $\cM^{i-1}$, which are defined in terms of the instanton
corrected Yukawa couplings $y_{abc}$ as
$$
y_{abc} = y^0_{abc}+\sum_{d_1,\dots,d_{h_{11}}}
{n^r_{d_1,\dots,d_{h_{11}}}d_ad_bd_c\over 1-\prod_{n=1}^{h_{11}}
q_n^{d_n}}\prod_{n=1}^{h_{11}}q_n^{d_n}
$$
\def\ni{n^r_{d_1,\dots,d_{h_{11}}}}
\noindent
where the $\ni$ is the virtual fundamental class of the moduli 
space of rational curves of multidegree $d_1,\dots,d_{h_{11}}$.
Such a relation between the instanton numbers are of course a special
property of the type of singularity we consider and will not be
present in other types of transitions proceeding e.g. through 
non-canonical singularities.

In a way similar to that  the special geometry of the vector moduli space 
restricts the relations between the coordinates on $\cM^i_V$
and $\cM_V^{i-1}$, the quaternionic structure of the hyper moduli 
space of the type IIB theory compactified on the same 
pair of manifold implies simple relations between the
coordinates on the hyper multiplet moduli spaces, $\xi_n$.
They are related to the algebraic moduli $z_n$ by rational
functions  which  in turn depend on the special representation
of the Calabi--Yau manifolds. In particular there are in general
different reflexive polyhedra describing the same Calabi--Yau space,
however in different algebraic coordinates related by rational
transformations. It is convenient to choose a preferred representation
in which the relation between the algebraic moduli on $\cM^i$ and
$\cM^{i-1}$ becomes particularly simple:
$$
z_n^{i-1} = \prod_m (z_m^{i})^{\delta_m^n}
$$
{}From the definition of the algebraic coordinates this relation translates to 
linear relations between the Mori vectors $l_n^i,\ l_n^{i-1}$,  generating
the
dual of the K\"ahler cone. As a consequence the smaller dual polyhedron
$\Delta^\star_{i-1}$ is obtained from the larger one $\Delta_i^\star$ 
by omitting one of the vertices\foot{This kind of transition was originally 
studied in~\ref\bkk{P. Berglund, S. Katz and 
A. Klemm, \npb456(1995)153,  hep-th/9506091.}\ and more recently in~\ref\cf{P. 
Candelas and A. Font, {\sl Duality Between Webs of Heterotic and Type 
II Vacua}, hep-th/9603170.}.}. 

A new feature of the corresponding transition in the moduli space
is that,
rather than beeing located at the zero locus of the
principal discriminant or a
restricted discriminant,
it can take place at one of the boundary divisors $z_n=0$.
Such transitions where first discussed in~\bkk; indeed this is the
situation in the examples described in  
section~3~\foot{A model with similar properties has been observed
in \bkk.}.  Similar examples have been considered 
independently~\ref\glass{D.R. Morrison, {\sl Through the looking glass},
to appear.}.

In the case when the Calabi--Yau is a K3 
fibration~\ref\oguiso{K. Oguiso, Internat. J. Math. 
{\bf 4} (1993) 439.}, the vector moduli space of the pair of 
Calabi--Yau manifolds can be mapped to that of the dual 
heterotic theory, the map again being linear by the above 
mentioned argument. Of course this relation will depend on 
the special compactification under consideration and 
will be discussed further in the examples.

\newsec{Global description of the Calabi-Yau spaces and examples}

Given a four-dimensional N=2
heterotic vacuum it is far from trivial to find the dual type IIA model.
Although there has been some progress recently as far as understanding the
perturbative gauge structure on the type IIA 
side~\AspinGauge\ref\hs{B. Hunt and R. Schimmrigk, {\sl Heterotic 
Gauge Structure of Type II K3 Fibrations}, hep-th/9512138.}\cf\ there 
is still some guess work to be done in terms
of finding a Calabi-Yau manifold that will fit the bill. In fact, recent 
work~\ref\bkkm{P. Berglund, S. Katz, A. Klemm and P. Mayr, 
{\sl Type II Strings on
Calabi-Yau Manifolds and String Duality}, in preparation.} 
indicates that there will in general exist more than one type IIA
theory which agrees with the perturbative heterotic vacuum under consideration.
{}From now on, we will assume that there is a set of models which to lowest order
corresponds to their counterparts in the  heterotic chain, i.e.\  the number of
vector multiplets and singlet hyper multiplets agree, and the models are
all K3-fibrations.

\subsec{A chain of connected Calabi--Yau manifolds}
We will here consider an example first discussed by Aldazabal et. al. \afiq. 
Let us first briefly review the structure of 
the chain of heterotic vacua with the least number of vector multiplets,
as it is the most suitable for the consideration in this paper; 
for more details see~\afiq .
Starting from a $Z_6$ orbifold with a specific embedding in 
$E_8\times E_8$ it is possible
to Higgs away most of the gauge group, leaving just
$SU(5)\times U(1)^4$~\foot{An alternate way of constructing this chain of 
models starts from the asymmetric instanton embedding $(10,14)$ which breaks 
$E_8\times E_8$ to $E_7\times E_7$, completely Higgsing one of the $E_7$ 
and the second $E_7$ to $SU(5)$\ref\afiqII{G. Aldazabal , A. Font , 
L.E. Ibanez , F. Quevedo, {\sl Heterotic/Heterotic Duality in D=6,4},
hep-th/9602097.}.}. 
The spectrum consists of 4 ${\bf 10}$, 22 ${\bf
5}$ and 118 ${\bf 1}$ hyper multiplets. In fact one can further Higgs
the $SU(5)$, giving a set of models with the following spectra. 
(In what
follows we will refer to the models by the number of vector multiplets and the
number of singlet hyper multiplets.)

\vskip10pt
{\vbox{\ninepoint{
$$\vbox{
{\offinterlineskip\tabskip=0pt
\halign{
\strut\vrule#&~#~&\hskip-6pt\vrule#&\hfil$ # 
$~&\vrule#&\hfil$ # $~&\vrule#&\hfil$ # $~&\vrule#\cr
\noalign{\hrule}
&$(N_v+1,N_h)$ 
&&{\rm gauge\  group }
&&{\rm spectrum} 
&&{\rm Calabi-Yau}
&\cr
\noalign{\hrule}
&$(8,118)$
&& SU(5)\times U(1)^4
&& $4\ {\bf 10}+22\ {\bf 5}+118\ {\bf 1}$
&& (\IP^5(1,1,2,5,7,9)[14 , 11])^{7,117}_{-220}& \cr
&$(7,139)$
&& SU(4)\times U(1)^4 
&& $4\ {\bf 6}+24\ {\bf 4}+139\ {\bf 1}$
&& (\IP^5(1,1,2,6,8,10)[16 ,12])^{6,138}_{-264}&\cr
&$(6,162)$
&& SU(3)\times U(1)^4
&& $30\ {\bf 3}+162\ {\bf 1}$
&& (\IP^4(1,1,2,6,8)[18])^{5,161}_{-312}&\cr
&$(5,191)$
&& SU(2)\times U(1)^4
&& $28\ {\bf 2}+191\ {\bf 1}$
&& (\IP^4(1,1,2,6,10)[20])^{4,190}_{-372}&\cr
&$(4,244)$
&& U(1)^4
&& $244\ {\bf 1}$
&& (\IP^4(1,1,2,8,12)[24])^{3,243}_{-480}&\cr
\noalign{\hrule}}}}
$$
\vskip0pt
\noindent{\bf Table 3.1:} Chain of heterotic - type IIA duals  
}
\vskip10pt}}

We have analyzed the matter structure and transitions for most of these
models.  Rather than give this analysis for the hypersurfaces and
complete intersections in weighted projective spaces here, 
we will find it more convenient to report on the transition 
using Batyrev's construction of Calabi-Yau mirror pairs 
$(X_n,X^*_n)$ as hypersurfaces\foot{We will indicate the number 
of vector multiplets of the models by the 
subscript $n$.} in more general  toric varieties 
$(P_{\Delta_n},P_{\Delta^*_n})$~\ref\batyrev{V. Batyrev, 
Duke Math. Journal 69 (1993) 349,
V. Batyrev, Journal Alg. Geom. 3 (1994) 493, 
V. Batyrev, {\sl Quantum Cohomology Rings of Toric Manifolds}, 
preprint 1992}. An exhaustive list of 
polyhedra $(\Delta_n,\Delta_n^*)$ for the models in~\afiq~can 
be found in~\cf. In general there will be several 
different candidates for $(X_n,X_n^*)$, differing presumably 
only by birational maps.

\subsec{Phases of the K\"ahler moduli space}

Given a singular ambient space $P_{\Delta}$, we have
in general many phases in the associated extended K\"ahler 
moduli space of the nonsingular space $\hat P_{\Delta}$.
They correspond to the different ways to resolve $P_{\Delta}$ 
and are defined by the different regular triangulations of 
the polytope $\Delta^*$. If $(\Delta,\Delta^*)$ are reflexive 
there is a canonical way\foot{Even if we fix the triangulation of
$\Delta^*$ this does not fix $(X_n,X^*_n)$ uniquely.
A simple counterexample with a non-toric phase is the 
$X_9(3,2,2,1,1)^{2,86}$ case discussed 
in~\bkk\ref\hly{S. Hosono, B.H.Lian, S.-T.Yau, 
{\sl GKZ Hypergeometric Systems and Applications to Mirror
                  Symmetry}, alg-geom/9511001},
\bkkm.} to embed a Calabi-hypersurface $(X,X^*)$ in 
$(\hat P_{\Delta},\hat P_{\Delta^*})$~\batyrev.
Among the phases of $\hat P_{\Delta}$ are the ones that give 
rise to Calabi-Yau varieties, when the K\"ahler classes are 
restricted to $X$. 
They correspond to triangulations\foot{We have
used PUNTOS~\ref\puntos{J.~DeLoera, PUNTOS,
ftp.geom.umn.edu/pub/software} to find the triangulations} involving 
all points of $\Delta^*$ on dimension $0,1,2$-faces.  
Even restricting to phases which correspond to manifolds which 
are K3-fibrations does not in general narrow down the choice 
to a unique model. Furthermore depending on the particular 
situation at hand, it may be the case that there exists more 
than one type IIA vacuum to a given heterotic theory. 
Finally, there is a technical problem in finding the true Calabi-Yau
phases. Frequently the K\"ahler cones of $\hat P_{\Delta}$ 
are narrower than the K\"ahler cone of $X$, 
because the former are bounded by curves in $\hat P_{\Delta}$ 
which vanish on $X$. We describe in Appendix A how to 
deal with this situation.

\subsec{The vector moduli space of our examples}

We will now use mirror symmetry and toric geometry to investigate 
the vector moduli space for the models in the chain described 
in the previous section. Candidates of type II models were constructed
as chains of nested polyhedra $\Delta^*_3\subset 
\ldots \subset\Delta^*_{10}$ ($\Delta_3 \supset\ldots 
\supset \Delta_{10}$)  in~\cf.
For simplicity, we now turn to studying the extremal transitions 
connecting the three
models with the fewest number of vector multiplets
discussed above. By extremal we refer to a
general transition obtained by contracting curves corresponding to
edges of the Mori cone, and then deforming the resulting singular
Calabi-Yau threefold to get a smooth Calabi-Yau manifold.  As will be shown 
this transition is not necessarily of the simple conifold 
type~\ref\AspinExtreme{P. Aspinwall, {\sl An N=2 Dual Pair and a 
Phase Transition}, hep-th/9510142.}.

We first have to calculate one valid Mori cone for each of the 
models\foot{We content ourselves with those phases arising from
the triangulations of $\Delta^*$.}. Let us therefore start by considering
$X_5$; as we will show, $X_4$ and $X_3$ can then be obtained by taking a 
particular limit in the K\"ahler moduli space of $X_5$.
Inside the  polyhedron $\Delta^*_5$ one has the following 
relevant points (inside dimension $0,1,2,4$-faces)~\cf 
\eqn\dualverts{\eqalign{
&
\nu^0=(0,0,0,0),\nu^1=(-1, 0, 2, 3),
\nu^2=(0, 0, -1, 0),\nu^3=(0, 0, 0, -1), 
\nu^4=(0, 0, 2, 3),\cr &\nu^5=(0, 1, 2, 3),\nu_6=(1, 2, 2, 3),
\nu^7=(0, -1, 2, 3), \nu^8=(0,-1,1,2), \nu^9=(0,-1,1,1).}}

{}The Mori cone can be found by repeated application of the 
procedure described in Appendix A. This leads to the 
following set of Mori generators
\eqn\moriIII{
\eqalign{&
l^{(1)}=(-1;0, 0, 0, 1, 0, 0, -2, 1, 1), 
l^{(2)}=(0 ;1, 0, 0, 0, -2, 1, 0, 0, 0),
\cr &
l^{(3)}=(0 ;0, 0, 0, -2, 1, 0, 1, 0, 0), 
l^{(4)}=(-1;0, 1, 0, 0, 0, 0, 1, -2, 1),
\cr &
l^{(5)}=(-1;0, 0, 1, 0, 0, 0, 0, 1, -1)}}

The mirror manifold of $X_5$ is given by the Laurent polynomial~\batyrev~
\eqn\laurent{P=\sum_{i=0}^9 \prod_{j=1}^4 a_i X_i^{\nu^i_j}=0}
in $P_{\Delta^*_5}$. A crucial insight \batyrev,\ref\morrison{D.\ R.\ Morrison, 
hep-th 9311049;
D.\ R.\ Morrison, P. Aspinwall, 
B.R. Greene, {\sl The monomial divisor mirror map},
hep-th/9407087}\ref\hkty{S. Hosono, A. Klemm, 
S. Theisen and S.-T. Yau,
\cmp167(1995)301, \npb433(1995)501.}, 
that the large complex structure variables of the mirror 
$X^*_5$ are defined by 
the Mori cone of $X$, specifically for $X_5$ above: 
\eqn\lcsv{w_i=(-1)^l \prod_{j=1}^9 a_j^{l^{(i)}}.}
By the construction of~\cf\ the models $X_4$, $X_3$ are given
as hypersurfaces in toric varieties whose dual polytopes, $\Delta^*_{4,3}$
respectively are obtained by deleting the point $\nu_9$ or points
$(\nu_8,\nu_9)$ from $\Delta_5^*$; this corresponds to the 
restriction of the moduli space of $X_5$ to $a_9=0$ and $a_9=a_8=0$ 
respectively.
  
{\vbox{\ninepoint{
$$\vbox{
{\offinterlineskip\tabskip=0pt
\halign{
\strut\vrule#&~#~&\hskip-6pt\vrule#&
\hfil$#$~&\vrule#&\hfil$#$~&\vrule#
&\hfil$#$~&\vrule#\cr
\noalign{\hrule}
&
&&\ X_3 
&&\ X_4
&&\ X_5 &\cr
\noalign{\hrule}
&$P$ 
&&a_8=a_9=0
&&a_8=0 
&& &\cr
\noalign{\hrule}
&$\Delta^*$ 
&&{\rm conv}(\nu_1,\ldots,\nu_7)
&&{\rm conv}(\nu_1,\ldots,\nu_8) 
&&{\rm conv}(\nu_1,\ldots,\nu_9)&\cr
\noalign{\hrule}
&$h^{1,1}$
&&3(0) 
&&4(0)
&&5(0)&\cr
\noalign{\hrule}
&$h^{2,1}$
&&243(1)
&&190(1)
&&161(1)&\cr
\noalign{\hrule}
&$z_1$
&& {w_1 w_4^2 w_5^3}
&& w_1 w_5 
&& w_1                  
&\cr
&$z_2$
&&w_2
&&w_2
&&w_2                    
&\cr
&$z_3$
&&w_3 
&&w_3
&&w_3                     
&\cr
&$z_4$
&&- 
&&w_4 w_5
&&w_4                     
&\cr
&$z_5$
&&- 
&&-
&&w_5                     
&\cr
\noalign{\hrule}}}}
$$
}}
\vskip-10pt
\noindent{\bf Table 3.2:} {\ninepoint{
The Calabi-Yau manifolds which correspond to reflexive polyhedra 
inside  $\Delta^*_5$. The polyhedra are 
specified as convex hulls of the points given in~\dualverts.
Furthermore, we list the number of K\"ahler $h^{1,1}$ and complex 
structure deformations $h^{2,1}$ of $X_i$ (the number of 
non-algebraic deformations is indicated in parentheses) as well as 
the vanishing coefficients in the Laurent polynomial~\laurent~and the 
canonical large complex structure 
coordinates $z_k$ of $X_i^*$, which are related to the 
Mori cones by \lcsv.
}}}
\vskip10pt

Using toric geometry one can calculate the classical intersections 
corresponding to the K\"ahler classes, $J_i$, which are dual
to the Mori generators~\moriIII, 

\eqn\ringIII{
\eqalign{
& 8J_1^3 + 2J_1^2J_2 + 4J_1^2J_3 + J_1J_2J_3 + 
   2J_1J_3^2 + \cr
& 16J_1^2J_4 + 4J_1J_2J_4 + 8J_1J_3J_4 + 
   2J_2J_3J_4 + 4J_3^2J_4 + 24J_1J_4^2 + 6J_2J_4^2 + 
   12J_3J_4^2 + 36J_4^3 +\cr
& 48J_5^2J_1+24J_5J_1^2+36J_5J_4J_1+6J_5J_2J_1+12J_5J_3J_1+
6J_5J_3^2+18J_5J_4J_3+\cr
& 3J_5J_3J_2+9J_5J_4J_2+12J_5^2J_2+
54J_5J_4^2+24J_5^2J_3+72J_5^2J_4+96J_5^3}}
as well as the evaluation of the Chern class on the $(1,1)$ forms 
$J_i$,
\eqn\cherntwo{
c_2(J_1)=92,\quad c_2(J_2)=24,\quad c_2(J_3)=48,\quad c_2(J_4)=132, \quad c_2(J_5)=168
}

Obviously the intersection numbers of $X_4$ and $X_3$ are simply given
from these by the restriction to the first four respectively three
K\"ahler classes. For further studying the transition we also need 
the Gromov-Witten invariant for the rational curves. These
are obtained from the solutions of the Picard-Fuchs equations
using the mirror hypothesis and listed in Appendix B.

\subsec{Local geometry and summation of the instanton corrections}

Let us begin with $X_3$, and identify the associated
toric variety $P_3$.~\foot{This model has been studied 
earlier~\ref\hkty,
in particular within the context of $N=2$ string duality in four
dimensions~\KaVa\ref\klm{A. Klemm, W. Lerche, P. Mayr, as cited
in \ogen }\AspinGauge.}\foot{In this section, we will frequently 
perform intersection calculations
in the Calabi-Yau manifolds $X_k$.  These calculations are often inferred
from the Mori cone; at times they may also be performed by
Schubert~\ref\schubert{S.~Katz and S.A.~Str\o mme.
Schubert: a Maple package for intersection theory,
ftp.math.okstate.edu/pub/schubert}.} 
A partial list on the number of rational curves
of low degree, including
those of importance for studying the transitions described in
this paper, can be found in table~A.1.

We will now show that this model is primitive, i.e.\  it does not admit
a geometric transition to a model with fewer K\"ahler parameters. Let
us therefore discuss the edges of the Mori cone one at a time, to
determine whether their contraction admits a birational smooth
deformation. We start by studying the 
first edge of the Mori cone.
This edge describes curves contained in an elliptic fibration over
a surface, so the contraction of these curves is not a birational
map (note that $n_{1,0,0}=480$, the negative
of the Euler characteristic, as explained in~\bkk).

For our purposes, it is best to think of $F_2$ as a complete
toric variety with edges $(-1,-2),(1,0),(0,1),(0,-1)$.  Its Mori cone
is given by $(-2,0,1,0,1),(0,1,-2,1,0)$.
As such, it can be
thought of as $\IC^4-(\{x_1=x_6=0\}\cup\{x_5=x_7=0\})$ (in terms of the
$x_i$ of $P_3$), modulo the 
$(\IC^*)^2$ identification 
\eqn\ident{(x_1,x_5,x_6,x_7)\sim(t x_1, s t^{-2} x_5, t x_6, s x_7).}
The hyperplane class of $F_2$ is the 
toric divisor $x_7=0$, and will be denoted by $H$.  We can also think of
$F_2$ as the minimal desingularization of $\IP(1,1,2)$, so it makes sense
to talk of the degree of a curve on $F_2$.  Note that a curve in the
class $dH$ has degree~$2d$.  In passing, we note that the
exceptional divisor of this blowup, $x_5=0$, is the section of 
self-intersection $-2$.

Returning to the first edge of the Mori cone, 
we note that the curve is contracted by 
$D_1,D_5,D_6,D_7$.\foot{Throughout our discussion, 
we will denote by $D_k$ the restriction to $X_i$ of the toric divisor 
with equation $x_k=0$, when the model under discussion is clear from context.}
The relations $D_1\cdot D_6=0$ and $D_5\cdot D_7=0$, together with the
$\IC^*$ actions defining the toric variety, show that there is a map
$X_3\to F_2$.  The fibers are elliptic curves with typical equation
$x_3^2+x_2^3+x_2f_{16}+f_{24}=0$, where the $f_i$ have degree $i$ in the
variables $x_1,x_6,x_7$ of $\IP(1,1,2)$, where $x_7$ is the variable of
degree~2.

The divisor $D_5$ describes a ruled surface over an elliptic curve.
This is the second edge of the Mori cone.
In this situation, the Gromov-Witten invariant is 
$2g-2=0$~\ref\cdfkm{P. Candelas, X. de la Ossa, A. Font,
S. Katz and D. Morrison, \npb{416}(1994)481, hep-th/9308083.}.  There
is no extremal transition in this case, although there is an $SU(2)$
gauge symmetry that is broken after a non-polynomial 
deformation~\kmp\foot{In~\ref\mvII{D. R. Morrison and C. Vafa,
{\sl Compactifications of F-Theory on Calabi--Yau Threefolds},
hep-th/96002114.}\ref\ago{P. S. Aspinwall and Mark Gross, 
{\sl Heterotic-Heterotic String Duality and Multiple K3 Fibrations}, 
hep-th/9602118.}
 it was shown that turning on the 
non-polynomial deformation connects the moduli space for $X_3$ defined as 
an elliptic fibration over $F_2$ (our case) with that of an elliptic fibration
over $F_0$.}.

Now we turn our attention to the divisor $D_4$.
The $K3$ fibration defined by $(x_1,x_6)$ restricts to $D_4$ to describe
$D_4$ as a ruled surface over a genus~0 curve; the Gromov-Witten invariant is
$n_{0,0,1}=2g-2=-2$, and there is no transition.

In summary, the Mori cone of $X_3$ coincides with that of $P_3$, and
this model is primitive, i.e.\ \ does not admit a geometric
transition to a model with fewer K\"ahler parameters.\foot{Recent relevant
geometric results about primitive Calabi-Yau threefolds have been given
in~\ref\GrossPrim{M. Gross, {\sl Primitive Calabi-Yau 
Threefolds}, alg-geom/9512002.}}  This
checks against the heterotic side, where the model with
$(n_H,n_V)=(244,4)$ is at the bottom of the chain.

We next turn to the 4~parameter model $X_4$.
Some of the instanton numbers for this model appear in table~A.2.
We will see that the fourth edge of the Mori cone is represented by
a conic bundle containing~28 line pairs, and that after contraction,
there is a transition to $X_3$. 
{}From this transition, the Mori cone of $X_3$ is the quotient of the
Mori cone of $X_4$ after modding out by the edge $(0,0,0,1)$.  It
remains to match up the edges from the above geometry.  The edges $(0,1,0,0)$
and $(0,1,0)$ correspond to the ruled surface over the elliptic curve, so
are to be identified.  The elliptic fibration identifies $(1,0,0,2)$ with
$(1,0,0)$; or equivalently identifies $(1,0,0,0)$ with $(1,0,0)$ due to
the quotient.  Finally, the remaining edges $(0,0,1,0)$ and $(0,0,1)$ are
identified as ruled surfaces over rational curves.
{}From this, we infer the relation
\eqn\relone{n_{a,b,c}=\sum_kn_{a,b,c,k}}
which checks against the instanton numbers that we have provided, for example
$n_{1,0,0}=-2+56+372+56+-2=480$.


As we saw for $X_3$, the model $X_4$ also admits a map $\pi:X_4\to F_2$
defined
by $(x_1,x_5,x_6,x_7x_8)$.  The fibers have type $(1,0,0,2)$ and are again
elliptic.  To calculate this type, note that since $x_1=x_5=0$ defines a point
of $F_2$, the same equation defines an elliptic fiber of $X_4$.  We accordingly
calculate the intersection numbers $D_1\cdot D_5\cdot D_k$ for $1\le k\le 8$,
obtaining the Mori vector $(-6; 0, 2, 3, 1, 0, 0, 0, 0)$, where the $-6$
arises because the coordinates must sum to 0.  In our basis for the 
Mori cone given in~(A.5), this is just $(1,0,0,2)$.  Throughout this section,
other classes have been computed in this manner; the classes will be given
without further comment.
A curve $C$ of type $(0,0,0,1)$ is also contracted by $\pi$.
Since $C\cdot D_8=-1$, we see that $C$ is necessarily contained in $D_8$.
Since $C\cdot D_7=1$, there is a unique fiber of $D_7$, a curve of type
$(1,0,0,0)$ which meets $C$.  This curve is also contracted by $\pi$.
The fiber of $\pi$ containing both of these curves contains a third
component of type
$(0,0,0,1)$; thus there are two curves of type $(0,0,0,1)$ in the same
fiber.

Because of $C\subset D_8$, we restrict attention to $D_8$, which admits
a map to $\IP^1$ by restricting the $K3$ fibration defined by $(x_1,x_6)$.
After restricting to $D_8$ (hence putting $x_8=0$), the equation of $X_4$ becomes
$x_2^2 f_8 + x_3^2 + x_7^2 f_{16} + x_2x_3 f_4 + x_2x_7 f_{12}
+x_3x_7 f_8=0$
where the $f_i$ have degree~$i$ in the variables $x_1,x_6$ of
$\IP^1$.  We interpret the above as a family of conics in the $\IP^2$
with coordinates $(x_2,x_3,x_7)$, with $(x_1,x_6)\in\IP^1$ as a
parameter.  The discriminant of this family has degree~24.
Thus the general fiber of $D_8$ is a smooth conic of type $(0,0,0,2)$,
while there are 28 fibers where the conic splits into line pairs, each
of type $(0,0,0,1)$.  As a check, note that $n_{0,0,0,2}=-2$ and
$n_{0,0,0,1}=56$.

The transition to $X_3$ is found by writing down monomials in the variables
of $P_4$ which have intersection number 0 with $(0,0,0,1)$.  Choosing
them in the order $(x_1,x_2x_8,x_3x_8,x_4,x_5,x_6,x_7x_8)$, we see that
the assignment
$$(x_1,\ldots,x_8)\mapsto (x_1,x_2x_8,x_3x_8,x_4,x_5,x_6,x_7x_8)$$
defines a mapping from $X_4$ to $P_3$ which contracts the conic bundle,
and takes $X_4$ to a singular form of $X_3$.  The transition is produced
simply by deforming the equation of $X_3$.

As a final comment on this model, we observe that the class of the elliptic
fiber is $(1,0,0,2)$.  We have seen that the fiber of $D_7$ is of type
$(1,0,0,0)$, while the fiber of $D_8$ is of type $(0,0,0,2)$.  The
intersection of $D_7$ and $D_8$ meets either fiber in two points. Thus
$D_7\cup D_8$ is a fibration over $\IP^1$ whose general fiber is a union
of two $\IP^1$s intersecting in two points, while there are 28 special
fibers which form triangles of curves.  By itself, $D_8$ is contracted to
get $X_3$, and this is the case $N=2,\ M=28$ of the geometry described in
Section~2.2.  We accordingly expect to see 28 ${\bf 2}$ hyper multiplets
becoming massless at the transition, and that is in perfect agreement with
Table~3.1.

Finally, we turn to the 5~parameter model $X_5$.
A partial list on low degree instanton numbers, including
those of importance for studying the transitions described in
this paper can be found in table~A.3. 

The key to understanding this transition is the contraction of the curves of
type $(0,0,0,0,1)$.  From table~A.3 we have $n_{0,0,0,0,1}=30$.  
This class has Mori
vector $(-1,0, 0, 1, 0, 0, 0, 0, 1, -1)$, so we see that this curve is 
contained in $D_9=0$ and is contracted by
the divisors $D_1,D_2,D_4,D_5,D_6,D_7$.  We also note from the first
entry of the Mori generators that the equation
of $X_5$ is in the class $J_1+J_4+J_5$, where the $J_k$ are the
dual generators of the K\"ahler cone.  We accordingly write the 
equation of $X_5$ in the form
\eqn\xfive{x_3 f + x_8 g,}
where $f$ is a polynomial with cohomology class $J_1+J_4$ and $g$
is a polynomial with cohomology class $3J_4$.  A curve is contracted
by the divisors listed above if and only if $f=g=x_9=0$.  We calculate
$D_9\cdot (J_1+J_4)\cdot 3J_4=30$.  Thus $n_{0,0,0,0,1}=30$.
The transition to $X_4$ is now visible---$X_4$ is obtained from $X_5$ by
the map
$(x_1\ldots,x_9)\mapsto(x_1,x_2,x_3x_9,x_4,x_5,x_6,x_7,x_8x_9).$

The Mori cone of $X_4$ is thus the quotient of the Mori cone of $X_5$
by the vector $(0,0,0,0,1)$.  We now match up the other edges.
The edge $(0,1,0,0,0)$ is the fiber of a ruled surface over an elliptic
curve, so corresponds to $(0,1,0,0)$.  There is again an elliptic fibration
over $\IP(1,1,2)$; we calculate that the fiber has class $(1,0,0,2,3)$, which
is equivalent to $(1,0,0,2,0)$ under the quotient.  This must match
with the elliptic class $(1,0,0,2)$ of $X_4$.  This implies that
$(0,0,0,1,0)$ corresponds to $(0,0,0,1)$, and $(1,0,0,0,0)$ corresponds
to $(1,0,0,0)$.  The remaining edges $(0,0,1,0,0)$ and $(0,0,1,0)$ are
therefore related; they are fibers of ruled surfaces over rational curves.

This gives the formula
\eqn\reltwo{n_{a,b,c,d}=\sum_k n_{a,c,b,d,k}}
which checks against the instanton numbers that we have provided, for example
$n_{1,0,0,0}=-2+30+30-2=56$.

As a final comment on this model, we observe that the class of the elliptic
fiber is $(1,0,0,2,3)$.  We have seen that the fiber of $D_7$ is of type
$(1,0,0,0,0)$, while the fiber of $D_8$ is of type $(0,0,0,1,0)$.  We
can also calculate that the fiber of $D_9$ has class $(0,0,0,1,3)$.
The curves of type $(0,0,0,0,1)$ are contained in degenerate fibers of $D_9$;
unlike the $X_4$ situation, the other component of this fiber is of a
different type $(0,0,0,1,2)$.  The $(0,0,0,0,1)$ curve meets $D_8$ but
not $D_7$, while the $(0,0,0,1,2)$ curve meets $D_7$ but not $D_8$.
Thus $D_7\cup D_8\cup D_9$ is a degenerate elliptic fibration over
$\IP^1$ whose general fiber is a triangle.  There are 30~special fibers
where there is an extra component, and the fiber is a square of curves.
Together, $D_8\cup D_9$ contract to give $X_3$ (note that if we contract
$(0,0,0,0,1)$ together with $D_8$ to get $X_3$, then we are contracting
$(0,0,0,1,0)$, hence the entire fibration $D_9$ with class
$(0,0,0,1,0)+3(0,0,0,0,1)$).  This is the case $N=3,\ M=30$ 
of the geometry described in Section~2.2.  We accordingly expect to see 
30 ${\bf 3}$ hyper multiplets
becoming massless at the transition, and that is in perfect agreement with
Table~3.1.

We can also explicitly see for the $X_5$ to $X_4$ transition
how the sets of three curves change to sets of two curves as we go to $X_4$.
After contracting $(0,0,0,0,1)$ to get to $X_4$, the two curves $(0,0,0,1,0)$
and $(0,0,0,1,2)$ pair up to become 30 pairs of $(0,0,0,1)$ curves. 
In fact, it can be shown that the curve $(0,0,0,1,2)$ is the ``partner'' 
of $(0,0,0,0,1)$ under the conic bundle. As the
conic bundle gets smoothed out by the deformation
process, there are fewer lines pairs left. 

We expect similar phenomena to arise for $X_6$ and $X_7$.  For example,
using the $\IP^5(1,1,2,6,8,10)[16,12]$ model for $X_6$, we have checked
that the transition to the $\IP^4(1,1,2,6,8)[18]$ model for $X_5$
occurs by projection on the first~5 coordinates, and there are 
$M=24$ exceptional curves lying on a birationally ruled surface.

\subsec{Physical interpretation of the examples}
Let us now try to understand the above described transitions in a physical 
context. We start with the $X_4$ model. At the codimension one surface 
where the conic bundle is contracted
we have a singular $\IP^1$ of type $A_1$ with 28 double points of type $A_2$.
As we will now argue, this corresponds to an infrared free theory $SU(2)$ 
gauge theory with 28 
${\bf 2}$. First, it has been 
shown in~\km~\kmp\ that a $\IP^1$ bundle over a 
curve with 
singularity type $A_1$ gives an enhanced 
$SU(2)$ gauge
symmetry in the type IIA string theory.  If the base curve of the
family is rational as in our case, it is also shown that there is
no new matter \kmp.
What we find here is that the contraction of the 
isolated curves, corresponding to solitonic  2-branes 
wrapped around the curves becoming massless, gives rise 
to non-abelian charged matter. The non-abelian charges arise
from the fact that these isolated curves originate from the
same conic bundle as does the continuous family of rational curves 
leading to the non-abelian gauge bosons. This is the non-abelian 
generalization of the conifold singularity.
As the $SU(2)$ is Higgsed, the rank of the gauge group is reduced 
by one, while the number of hyper multiplets increase by 53. This is seen 
very nicely from the expression of  the instanton numbers~\relone. 
Note how the 
fact that $n_{0,0,0,2}=-2$ fits with losing two of the hyper multiplets 
as the $W^{\pm}$
of the $SU(2)$ becoming massive. There is of course as usual one 
hyper multiplet
 which is ``eaten'' as the $U(1)$ gauge boson of the Cartan subalgebra of
$SU(2)$ become massive. 
Note how this exactly matches the heterotic description.  
At the transition from $(191,5)$ to $(244,4)$, the gauge
group is $SU(2)\times U(1)^4$, with 28 ${\bf 2}$ hyper multiplets
under the $SU(2)$.  The transition occurs by Higgsing the $SU(2)$.

In fact we can give a quite explicit map of the transition of the
type II theory to that on the heterotic side by analyzing the
physical quantities such as the mirror maps, discriminants and periods
which determine the $N=2$ effective action. The relation between the 
Mori generators
\eqn\morirel{
l_1^{(3)}=l_1^{(4)}+2\ l_4^{(4)},\ l_2^{(3)}=l_2^{(4)},\ l_3^{(3)}=l_3^{(4)}
}
implies the following relations between the special and algebraic coordinates,
in an obvious notation:
\eqn\coordrel{\eqalign{
z_1^{(3)}&= z_1\ z_4^2,\ z_2^{(3)}= z_2,\ z_3^{(3)}= z_3\cr
t_1^{(3)}&=t_1+2\ t_4,\ t_2^{(3)}=t_2,\ t_3^{(3)}=t_3 \ .}
}
{}From the mirror maps we find 
$$
z_4 \sim {1 \over (1-q_4)^2},\ z_1\sim(1-q_4)^4
$$
that is the transition takes indeed place at the boundary divisor 
$z_1=1/z_4=0$. In this limit the data of $X_4$ such as the periods and
discriminants reduce to those of $X_3$. It is instructive to see the
connection to the heterotic moduli. On the heterotic side, which is
realized in terms of a simple  orbifold construction of $K3$~\afiq\ 
one has in addition to the dilaton $S$, the moduli of the torus $T,U$ 
a Wilson line $B$. On these moduli acts the perturbative T-duality
group which in a similar model has been determined to be~\ref\ms{
P. Mayr and S. Stieberger, \plb{355}  (1995) 107;
G.\ Lopes Cardoso, D.\ L\"ust and T. Mohaupt, \npb432 (1994) 68}:
\eqn\duals{\eqalign{
&T\to T+1,\ U\to U+1,\ B\to B+1,\ B \to -B,\ T \leftrightarrow U;\cr
&B\to B-U,\ T\to T+U-2B,\ U\to U,
}}
together with the generalization of the inversion element 
$T\to-1/T$ which is however realized in a less obvious way on the
physical expressions. We expect a similar modular group realized in
the present model, possibly up to some coefficients which depend
on the details of the compactification lattice.
Indeed, matching \duals\ to the symmetries realized on the
physical couplings of the Calabi--Yau compactification,
we find the identifications
$$
t_1=T,\ t_2=S-T,\ t_3=U-T,\ t_4=B+U
$$
where $B$ has in fact period $2\ U$ rather than $U$. As a check on
our physical picture we note that the Weyl symmetry element
of the $SU(2)$ subgroup  of the $E_8$ factor is not 
corrected by non-perturbative string effects contrary to 
the mirror symmetry of the heterotic torus, as expected.

The situation for $X_5$ is similar but as the rank is larger there is now
room for more interesting phenomena. As for $X_4$ we can obtain an
$SU(2)$ by shrinking down the divisor $D_8$. As this is not the
conic bundle  there are no 
degenerate fibers, i.e. we have a family of curves parameterized by $\IP^1$ 
which is reflected in $n_{0,0,0,1,0}=-2$. 
Hence, there is no matter, and the unbroken 
$SU(2)$ is
present only in the perturbative theory. This agrees with the general field 
theoretic picture, as discussed in section~2.1. When the rank of the gauge 
group is 2, there is an $SU(2)$ for $\phi_1=\phi_2\neq 0$, where the $\phi_i$
are the scalar vevs of the vector multiplets. If we in addition to 
contracting the instantons of degree $(0,0,0,1,0)$ also shrink those of type
$(0,0,0,0,1)$ we get a further enhancement, to $SU(3)$ as well as 30 massless
triplets. This can be seen as we are now forced to shrink down any combination
of instantons which have degree $(0,0,0,k,l)$. Among the non-zero entries we
find three components which all have $n_{(0,0,0,k,l)}=30$. Thus, 3 times 30
massless particles, forming 30 ${\bf 3}$s under $SU(3)$. As we Higgs the
$SU(3)$, the three components split into a set of two which gives the 28 
${\bf 2}$s of the remaining $SU(2)$ and 29 new singlet hyper multiplets. 
We have
then arrived at the $SU(2)$ point of $X_4$ discussed above. Once
more this agrees perfectly with the heterotic picture.
At the transition from $(162,6)$ to $(191,5)$, the gauge group is
$SU(3)\times U(1)^4$, with 30 ${\bf 3}$ hyper multiplets under the
$SU(3)$.  The transition occurs by breaking the $SU(3)$ to $SU(2)$ just as 
described above.

Finally, it is possible to avoid the enhanced gauge symmetry, and restrict to 
a codimension one surface where 30 isolated instantons of degree $(0,0,0,0,1)$
shrink to zero. This is the type IIA  analog of the conifold transition
in the type IIB string discussed by Greene, 
Morrison and Strominger~\ref\gms{B.R. Greene, 
D.R. Morrison, A. Strominger, {\sl Black Hole Condensation and the 
Unification of String Vacua}, hep-th/9504145.}.
Here, there are 29 flat directions among the 30 hyper multiplets 
which become
massless as the size of the instantons go to zero.  
Thus, after Higgsing we
are left with $U(1)^5$ and 29 new singlets.  This IIA transition description
was given in~\kmp, and applies as well to a similar transition occuring 
in~\bkk.

\newsec{Discussion and Conclusions}
In this paper we have given strong evidence for extremal transitions
between type II Calabi-Yau vacua, where the dual process in the
heterotic string corresponds to Higgsing\foot{In a recent paper, transitions
between type II vacua related to the dual $SO(32)$ heterotic string
have been discussed~\ref\ag{P. S. Aspinwall and M. Gross, {\sl The SO(32) 
Heterotic String on a K3 Surface}, hep-th/9605131.}.}. 
In particular, we have found evidence for a very nice correspondence 
between the appearance of enhanced $SU(N)$ gauge symmetries and  the 
corresponding matter structure on one hand, and the existence of a 
particular type of singularities in the Calabi-Yau manifold. The geometrical
structure in question is that of $N-2$ rationally ruled surfaces and a conic 
bundle with $M$ degenerate fibers. As the fibers, which are $\IP^1$s,
 shrink to zero, particles appearing as BPS-saturated states of 2-branes
wrapped around the $\IP^1$s become massless. The crucial point is the existence
of the degenerate fibers as they are the source of the massless matter transforming
in the fundamental of the relevant gauge group. The particular models 
considered in this paper are elliptically fibered Calabi-Yau manifolds. 
However, it seems as if the existence of the conic bundles is independent of
this fact. We thus believe that this scenario is more general, and we are
currently investigating such models\foot{It has become known to us that 
similar problems are to be discussed in a forthcoming paper by
Bershadsky et. al~\ref\bvw{Bershadsky et. al, {\sl Geometric Singularities
and Enhanced Gauge Symmetries}, to appear.}.}.

Finally, recall that the
transitions are taking place in the perturbative region of the
heterotic string, i.e.\  in the limit of large base in the K3-fibration in 
the type II theory. Thus, it still remains the possibility that there
exist other K3-fibrations with the same behaviour as the radius of
the $\IP^1$ becomes large~\bkkm.

\vskip20pt
\noindent
{\bf Acknowledgments}:
We thank Philip Candelas, Anamaria Font, Wolfgang Lerche, 
Bong Lian, Jan Louis, 
 Fernando Quevedo, S-S. Roan, Rolf Schimmrigk, Andy Strominger,
Bernd Sturmfels, Stefan Theisen and S-T. Yau for useful discussions
and in particular Dave Morrison and Ronen Plesser for explanations
on the relevant singularity structure.
P.B.\ would like to thank the Institute for Advanced Study where this 
work was iniated, and the Institute for Theoretical Physics at the University
of California at Santa Barbara and
the Department of Mathematics at Oklahoma State University
for hospitality during parts of this project.
P.B.\ was supported by the DOE grant DE-FG02-90ER40542 and by the 
National Science Foundation under grant No. PHY94-07194. S.K.\ was
supported by NSF grant DMS-9311386 and an NSA grant MDA904-96-1-0021.

\appendix{A}{The K\"ahler Cone: Toric Variety vs. Calabi-Yau Hypersurface}

We will now explain the relation between phases, although
different when thought of as toric varieties, in fact are identical when
restricted to the Calabi-Yau hypersurface. Let us assume that we have two
toric varietes, $P_I$ and $P_{II}$, which are related to each other by a flop, 
i.e.\  
a surface/curve, $C$, is blown down on $P_I$, and when passing through the 
wall of
the K\"ahler cone where $P_I$ and $P_{II}$ meet $C$ reemerges in $P_{II}$
 as a 
surface/curve, $\tilde C$. (In terms of the complex structure moduli space of
the mirror theory the flop is merely an analytic continuation beyond the
radius of convergence, corresponding to the walls of the K\"ahler cone.)
We are still, however, to restrict this process to that of the hypersurfaces
$X_I$ and $X_{II}$.
Indeed, if the restriction of $C$ and $\tilde C$ to $X_I$ and $X_{II}$
respectively is empty there is nothing to be flopped and $X_I$ is
isomorphic to $X_{II}$. We then have to consider the new K\"ahler cone 
as that of the union of the K\"ahler cones of the $P_{I,II}$. This process 
is repeated until we have a distinct set of inequivalent models. (In the
example we will consider we always find just one K3-fibration phase after
applying the above scheme.)

Let us now apply the above idea to that of toric variety $P_4$
{}From the dual polytope $\Delta_4^*$ one finds three Calabi-Yau 
phases which all are $K3$-fibrations. 
Their respective Mori generators are given by
\eqn\moriIIa{
\eqalign{&(0, 0, -1, 0, 1, 0, 0, -3, 3),
(0, 1, 0, 0, 0, -2, 1, 0, 0),\cr
& (0, 0, 0, 0, -2, 1, 0, 1, 0), 
(-2, 0, 1, 1, 0, 0, 0, 1, -1)}}
\eqn\moriIIb{
\eqalign{&(0, 1, 0, 0, 0, -2, 1, 0, 0), (-2, 0, 0, 1, 1, 0, 0, -2, 2),\cr
& (0, 0, 1, 0, -1, 0, 0, 3, -3), (0, 0, -1, 0, -5, 3, 0, 0, 3)}}
and 
\eqn\moriIIc{
\eqalign{&(0, 0, 1, 0, 5, -3, 0, 0, -3),
(0, 1, 0, 0, 0, -2, 1, 0, 0),\cr
&(0, 0, 0, 0, -2, 1, 0, 1, 0),
(-2, 0, 0, 1, 1, 0, 0, -2, 2). }}
The second toric variety is obtained from the third one by a birational
transformation which contracts the surface $x_2=x_4=0$ on the second toric 
variety and resolves the resulting singularity to the surface $x_5=x_8=0$
on the third toric variety. But on the second
model for $X_4$ we have $D_2\cdot D_4=0$, and on the second we
have $D_5\cdot D_8=0$.  This says that the birational tranformation does
not affect the hypersurface, which are therefore isomorphic.  So there
is really just one Calabi-Yau phase coming from the two toric varieties
described by~\moriIIb\ and \moriIIc. Therefore the K\"ahler cone  
of the hypersurface is the union of the two
K\"ahler cones.  Since the Mori cone is dual to the K\"ahler cone,
we conclude that the actual Mori cone is the intersection of the two
Mori cones~\moriIIb\ and \moriIIc, which we calculate to be
\eqn\moriIIbc{
\eqalign{&(0, 1, 0, 0, 0, -2, 1, 0, 0),(-2, 0, 0, 1, 1, 0, 0, -2, 2),\cr
&(0, 0, 0, 0, -2, 1, 0, 1, 0),(0, 0, 1, 0, -1, 0, 0, 3, -3). }}
However, this new toric variety is related to that of \moriIIa\ by a flop
as well; contracting $x_4=x_8=0$ in the above phase and then resolving the 
surface $x_2=x_7=0$ in phase $I$. However, just as in the previous case
$D_4\cdot D_8=0$ when restricted to the Calabi-Yau hypersurface 
in \moriIIbc\ and $D_2\cdot D_7=0$ on the hypersurface in phase $I$. Thus we 
are left with just one Calabi-Yau phase given as a hypersurface in a toric 
variety where the Mori cone is generated by
\eqn\moriIIabc{
\eqalign{&(-2, 0, 0, 1, 1, 0, 0, -2, 2),(0, 1, 0, 0, 0, -2, 1, 0, 0)\cr
&(0, 0, 0, 0, -2, 1, 0, 1, 0),(-2, 0, 1, 1, 0, 0, 0, 1, -1) . }}

\appendix{B}{The Gromov-Witten invariants for $X_{3,4,5}$}         
\vskip7pt
\vbox{
{\ninepoint{
$$
\vbox{\offinterlineskip\tabskip=0pt
\halign{\strut\vrule#
&~$#$~\hfil
&~\hfil$#$\quad\quad
&~$#$~\hfil
&\hfil~$#$\quad\quad
&~$#$~\hfil
&\hfil~$#$\quad\quad
&~$#$~\hfil
&\hfil~$#$\quad\quad
&\vrule#\cr
\noalign{\hrule}
&[0, 0, 1] & -2& [0, 1, 1] & -2&[0, 1, 2] &-4&[0, 1, 3]& -6&\cr
& [0, 1, 4] &  -8 & [0, 1, 5] &  -10 & [0, 2, 3] &  -6 & [0, 2, 4] &  -32&\cr
& [1, 0, 0] & \ph 480 & [1, 0, 1] & \ph 480 & [1, 1, 1] &  \ph 480 
& [1, 1, 2] &  \ph 1440 &\cr
& [1, 1, 3] &  \ph 2400& [1, 1, 4] &  \ph 3360& [1, 2, 3] &  \ph 2400
& [2, 0, 0] &  \ph 480 &\cr
& [2, 0, 2] &  \ph 480 & [2, 2, 2] &  \ph 480& [3, 0, 0] &  \ph 480
& [3, 0, 3] &  \ph 480&\cr
& [4, 0, 0] &  \ph 480& [5, 0, 0] &  \ph 480 & [6, 0, 0] &  \ph 480
& [0, 1, 0] &  \ph 0&\cr
\noalign{\hrule}}
\hrule}$$
\noindent
{\bf Table B.1}: A list of instanton numbers for rational curves
of degree $[a_1,a_2,a_3]$ on $X_3$.}}}
\bigskip

\vskip7pt
\vbox{
{\ninepoint{
$$
\vbox{\offinterlineskip\tabskip=0pt
\halign{\strut\vrule#
&~$#$~\hfil
&~\hfil$#$\quad\quad
&~$#$~\hfil
&\hfil~$#$\quad\quad
&~$#$~\hfil
&\hfil~$#$\quad\quad
&~$#$~\hfil
&\hfil~$#$\quad\quad
&\vrule#\cr
\noalign{\hrule}
&[0, 0, 0, 1] & \ph 56& [0, 0, 0, 2] & -2&[0, 0, 0, 3]&\ph 0&
[0, 0, 1, 0]& -2&\cr
& [0, 1, 0, 0] &  \ph 0 &  [0, 1, 1, 0]&  -2 & [0, 1, 2, 0] &  -4 
&  [0, 1, 3, 0]&  -6&\cr
& [0, 1, 4, 0] & -8 & [0, 1, 5, 0] & -10 & [0, 2, 3, 0]  &  -6 
& [0, 2, 4, 0]  &  -32&\cr
& [0, 2, 5, 0] &  -110&[0, 2, 6, 0]   &-288 & [0, 3, 4, 0] &  -8
& [0, 3, 4, 0] &  -8 &\cr
& [0, 3, 5, 0] & -110  & [1, 0, 0, 0] &  -2& [1, 0, 0, 1] & 56 
& [1, 0, 0, 2] &  372&\cr
& [1, 0, 0, 3]  & 56 & [1, 0, 0, 4] & -2  &  &  
&  &  \ph 0&\cr
\noalign{\hrule}}
\hrule}$$
\noindent
{\bf Table B.2}: A list of instanton numbers for rational curves
of degree $[a_1,a_2,a_3,a_4]$ on $X_4$.}}}

\bigskip
\vskip7pt
\vbox{
{\ninepoint{
$$
\vbox{\offinterlineskip\tabskip=0pt
\halign{\strut\vrule#
&~$#$~\hfil
&~\hfil$#$\quad\quad
&~$#$~\hfil
&\hfil~$#$\quad\quad
&~$#$~\hfil
&\hfil~$#$\quad\quad
&~$#$~\hfil
&\hfil~$#$\quad\quad
&\vrule#\cr
\noalign{\hrule}
& [0, 0, 0, 0, 1] & 30& [0, 0, 0, 0, 2] & 0&[0, 0, 0, 1, 0] &-2&
[0, 0, 0, 1, 1]& 30&\cr
&[0, 0, 0, 1, 2] &  30 & [0, 0, 0, 1, 3]  &  -2 & [0, 0, 1, 0, 0] &  -2 &
[0, 1, 0, 0, 0] &  0&\cr
& [0, 1, 1, 0, 0] & -2 & [0, 1, 2, 0, 0] &-4 & [0, 1, 3, 0, 0] &  -6 
& [0, 1, 4, 0, 0] &  -8 &\cr
& [0, 2, 3, 0, 0]& -6 & [0, 2, 4, 0, 0]  & -32  & [0, 2, 5, 0, 0] & 
-110& [1, 0, 0, 0, 0] & -2  &\cr
& [1, 0, 0, 1, 0] & -2 & [1, 0, 0, 1, 1]  & 30 & [1, 0, 0, 1, 2] &  30
& [1, 0, 0, 1, 3] & -2 &\cr
& [1, 0, 0, 2, 2] & 30 & [1, 0, 0, 2, 3] & 312  & [1, 0, 0, 2, 4] & 30 
& [1, 0, 1, 0, 0] & -2 &\cr
& [1, 0, 1, 1, 0] & -2 & [1, 0, 1, 1, 1] & 30  &[1, 0, 1, 1, 2]  & 30 
& [1, 0, 1, 1, 3] & -2 &\cr
\noalign{\hrule}}
\hrule}$$
\noindent
{\bf Table B.3}: A list of instanton numbers for rational curves
of degree $[a_1,a_2,a_3,a_4,a_5]$ on $X_5$.}}}\hb

\listrefs
\bye